# The Viscosity of Polyelectrolyte Solutions and its Dependence on their Persistence Length, Concentration and Solvent Quality

E. Mayoral[1], J. D. Hernández Velázquez[2] and A. Gama-Goicochea[2*]

[1]Instituto Nacional de Investigaciones Nucleares, Ocoyoacac 52750, Estado de México, Mexico

[2]División de Ingeniería Química y Bioquímica, Tecnológico de Estudios Superiores de Ecatepec, Ecatepec de Morelos 55210, Estado de México, Mexico

**ABSTRACT**

In this work, a comprehensive study about the influence on shear viscosity of polyelectrolyte concentration, persistence length, salt concentration and solvent quality is reported, using numerical simulations of confined solutions under stationary Poiseuille flow. Various scaling regimes for the viscosity are reproduced, both under good solvent and theta solvent conditions. The key role played by the electrostatic interactions in the viscosity is borne out when the ionic strength is varied. It is argued that these results are helpful for the understanding of viscosity scaling in entangled polyelectrolyte solutions for both rigid and flexible polyelectrolytes in different solvents, which is needed to perform intelligent design of new polyelectrolytes capable of fine tuning the viscosity in complex fluids.

---

[*] Corresponding author. Electronic mail: agama@alumni.stanford.edu

# I. INTRODUCTION

Polyelectrolytes (PEs) are charged polymers present in natural surroundings such as in biological macromolecules and are frequently present in many industrial processes, such as in the drilling fluids for enhanced oil recovery, drug delivery, drag reduction, and flocculants, among others.[1,2] PEs display typically different behavior from neutral polymers, particularly in regard to their colligative and transport properties, due to their electrostatic interactions among the charged groups along the chain. PE solutions have been the subject of extensive theoretical and experimental research, to seek for fundamental understanding of their equilibrium and transport properties. PE solution rheology is a complex issue;[2] on the one hand, the chain conformation within the electrostatic blob depends on solvent quality, that is, the thermodynamic interactions between the solvent and the polymer. However, it is also known that ions in the solution affect the PEs interactions.[3–13] It is known that the chain structure of PEs is rodlike up to the correlation length ($\xi$), while for length scales larger than $\xi$ it is flexible and can be described by a random walk of correlation blobs.[2,14] Numerous scaling regimes for the viscosity ($\eta$) and the characteristic relaxation time ($\tau$) for flexible PEs in salt-free solutions have been proposed.[2] In particular, when PE concentration exceeds the overlap concentration ($c^*$), the relative viscosity ($\eta_0$) is found to be proportional to the square root of concentration $c_p$ as ($\eta_0 \sim c_p^{1/2}$).[1,5]

Recent work has explored the influence of the flexibility of PEs on their rheology,[15] but a complete understanding of its effect on solution viscosity, critical PE concentration, and the role of ionic strength from the dilute to the entangled regime is still lacking. Also, a comprehensive examination of chain stiffness on the scaling of the viscosity of PEs has not been achieved. The scaling of the viscosity has been studied as a function of PE concentration,[15] yet the relationship between chain flexibility, PE concentration, and solvent quality has not been thoroughly investigated. Further research is essential to improve the understanding of viscosity in entangled PE solutions for both rigid and flexible chains.

The relation between the viscosity of a PE solution and the polymer concentration ($c_p$) follows power laws in the unentangled, semi-dilute regime that differ from those for neutral polymer solutions.[16–18] Classical theories predict the empirical Fuoss law $\eta \propto c_p^{1/2}$ for PE solutions with no added salt (nominally "salt-free"),[1,14,19] and $\eta \propto c_p^{5/4}$ when salt is added.[2,3] Recently, new power-law exponents have been measured for salt-free PE solutions, finding 0.68 for sodium carboxymethylcellulose (NaCMC)[20–22] and acrylamide-sodium-2-acrylamido-2-methylpropane-sulfonate (AM-NaAMPS),[23] and 0.91 for salt-free hyaluronic acid (HA)[24] and chitosan,[25] all of which have different flexibilities. Although scaling theories are known to be applicable to flexible PE, characteristics such as the polymer persistence length $\xi$ are not taken into account by those theories.[2,14,26] Numerical simulations are powerful tools to explore different scenarios and complement the predictions of analytical theories that have been proposed to explain experimental trends in properties of both polyelectrolytes and neutral polymers.[2,14,19,27–30]

Through molecular dynamics simulations, it has been found that the PE concentration plays a fundamental role in the viscosification of aqueous solutions.[31–33] Mintis and co-workers[31–33] successfully reproduced experiments of zero shear rate viscosity ($\eta_0$) in salt-free aqueous solutions, as a function of the PE concentration for poly(*N,N*-dimethylaminoethyl methacrylate) (PDMAEMA)[33] using atomistic simulations. Although atomic-scale molecular simulations are very accurate, the modeling of shear viscosity requires long computational times, especially for systems with long-range interactions such as the electrostatic interaction. A useful alternative is coarse-grained computer simulations, where the short-range, non-electrostatic interactions allow for the modeling of entangled PE complex solutions.[34] Here we report mesoscale numerical simulations of PEs in an aqueous solution under a Poiseuille flow for increasing stiffness of the PE chains and predict the viscosity. The influence of solvent quality on viscosity is studied as well as the net charge of the coarse-grained salt ions.

## II. MODELS AND METHODS

We perform a set of coarse-grained simulations using the well-known dissipative particle dynamics (DPD)[35–37] model to investigate the influence of PE concentration, persistence length, salt concentration and solvent quality on the viscosity of confined solutions under stationary Poiseuille flow. The DPD model has proven to be useful in the study of several systems under stationary[38–40] and oscillatory flows.[41,42] In the DPD model, the total force ($\mathbf{f}_i$) that any DPD particle experiences is the sum of three fundamental pairwise forces, namely dissipative $\mathbf{F}_{ij}^D$, random $\mathbf{F}_{ij}^R$ and conservative forces $\mathbf{F}_{ij}^C$ ($\mathbf{f}_i = \sum_{j \neq i}[\mathbf{F}_{ij}^D + \mathbf{F}_{ij}^R + \mathbf{F}_{ij}^C]$). These short-ranged forces are additive central forces that obey Newton's third law and conserve momentum. Here, the maximum strength constants of $\mathbf{F}_{ij}^D$ and $\mathbf{F}_{ij}^R$ are coupled through the fluctuation-dissipation theorem.[36] The conservative force $\mathbf{F}_{ij}^C = a_{ij}(1 - r_{ij}/r_c)\hat{\mathbf{r}}_{ij}$, is a soft, linearly decaying repulsive force that depends on the relative distance between particles ($r_{ij}$), where $\hat{\mathbf{r}}_{ij} = (\mathbf{r}_i - \mathbf{r}_j)/|r_i - r_j|$ is the unit position vector. The maximum strength of this force is given by the parameter $a_{ij}$ when the relative distance between particles is smaller than a cutoff radius ($r_{ij} < r_c$), and $\mathbf{F}_{ij}^C = 0$ when $r_{ij} \geq r_c$, with $r_c = 1$. The $a_{ij}$ parameters are obtained directly from the Flory-Huggins solubility parameter. Further details can be found elsewhere.[37] Table 1 displays the $a_{ij}$ parameters used in the systems modeled in this work. These values correspond to a coarse-graining degree equal to three, i.e., the volume of a DPD bead equals the volume of three water molecules.[37]

**Table 1**. Conservative DPD force parameters $a_{ij}$, used in this work. The listed values correspond to PEs under good solvent conditions. The values in parentheses correspond to values of the PEs in theta solvent. The nomenclature is as follows: S = solvent beads, PE = polyelectrolyte monomer beads, and C = counterion beads. Values are shown in reduced units.

| $a_{ij}$ | S | PE | C |
|---|---|---|---|
| S | 78.3 | 60.0 (78.3) | 60.0 (78.3) |
| PE |  | 78.3 | 79.3 (78.3) |
| C |  |  | 78.3 |

In addition to the three fundamental DPD forces, additional forces are needed to model chain-like structures (surfactants, polymers, PEs, etc.) to bond two and three consecutive DPD particles conforming the linear chain. Thus, we add two harmonic forces; one is a Hookean force based on the Kremer – Grest bead – spring model[43]

$$\mathbf{F}_{ij}^S = -k_S(r_{ij} - r_0)\hat{\mathbf{r}}_{ij}, \tag{1}$$

to simulate the bonding between two consecutive DPD particles, where $r_{ij}$ is the relative distance between the particles, $k_S$ the spring constant, and $r_0$ the equilibrium distance. For all our simulations, we set $k_s = 100$ and $r_0 = 0.7$. It has been proven that properties such as the excess pressure and the interfacial tension are insensitive to changes in the values of $k_s$ and $r_0$ chosen in in this work.[44] To modulate the flexibility of the linear chain, a three – body force is introduced, modeled as a harmonic potential acting between two consecutive linear bonds (three consecutive DPD particles):

$$F_{ijk}^A = -k_A(\theta_{ijk} - \theta_0) \qquad (2)$$

where, $\theta_{ijk}$ is the angle formed by three consecutive DPD particles, $k_A$ is the angular spring constant and $\theta_0$ are the equilibrium angle. For this three – body force, we set the equilibrium angle as $\theta_0 = 180°$, while the values of $k_A$ used in this work are presented in Table 2, as well as their corresponding values of the persistence length ($L_p$), obtained from the following equation[45]

$$L_p = -\frac{r_0}{\ln[L(k_A)]} \qquad (3)$$

where, $r_0$ is the bond length (see eq. (1)), and $L(k_A) = \coth(k_A) - (1/k_A)$ is Langevin's function and $k_A$ the angular spring constant (see eq. (2)). It is emphasized that here we model generic PEs, without mapping their chemical nature to a particular PE because our purpose is to study the viscosity scaling of PEs with increasing stiffness under different solvent conditions and PE concentration.

**Table 2**. Angular spring constant values ($k_A$, eq. (2)) with their equivalent value of persistence length ($L_p$, eq. (3)). $k_A$ values are represented in reduced units, while $L_p$ values are reported in *nm*.

| $k_A$ | 2.5 | 4 | 10 | 30 | 50 | 90 | 100 | 110 | 150 | 180 | 200 | 210 | 500 |
|---|---|---|---|---|---|---|---|---|---|---|---|---|---|
| $L_p$ | 0.93 | 1.58 | 4.29 | 13.24 | 22.38 | 40.47 | 44.99 | 49.52 | 67.60 | 81.17 | 91.21 | 94.74 | 225.87 |

To produce parabolic flow, the system is confined by effective surfaces. Confinement is implemented perpendicularly to the *z*-direction, using a unidirectional implicit force placed at the top ($+l_z/2$) and bottom ($-l_z/2$) sides of the simulation box. These effective walls act over all the DPD particles and are modeled as a short-ranged, linearly decaying forces

$$\boldsymbol{F}_i^W = \begin{cases} a_w(1 - z_i/z_c)\hat{\boldsymbol{z}}_i, & z_i < z_c \\ 0, & z_i \geq z_c \end{cases} \quad (4)$$

where, $a_w$ is the maximum repulsion strength that the walls exert of any particle, and $z_c$ is the cutoff length of the wall force. In this work, the walls are separated by a distance $D = l_z = 23r_c$, and we fixed the values of $a_w = 115$ and $z_c = 0.4r_c$.[34] Although, there is a recent study showing how to control the particle penetration at the walls, using explicit walls made up of DPD particles,[46] the implicit force law used here to simulate the wall is strong enough to prevent penetration of the surfaces. Artifacts introduced by the velocity profiles produced by the Poiseuille flow when Lees – Edwards periodic boundary conditions are used can be avoided with this setup. It allows one to calculate the shear viscosity of fluids as a function of shear rate, by calculating the stress tensor in response to the shear rate.[47] The Poiseuille flow velocity profile in the system is obtained by applying an additional constant force to every particle.[38] In this setup, a small constant external force is applied along the *x*-direction, $\mathbf{F}_x^P = 0.02\hat{\mathbf{x}}$, to all particles in the system. Zero velocity along the *z*-direction is applied to the solvent particles whose distance from either wall is $z \leq 0.15r_c$, to avoid slip.[48] This restriction is applied only to solvent beads, since imposing to PE chains would graft them to the surfaces. In Fig. 1(a), a schematic representation of the setup used in this work is presented, showing the PEs confined by two parallel walls (see eq. (4)) separated a distance $D = l_z = 23r_c$. An external constant force along the *x*-axis acts on all beads, producing a parabolic velocity profile typical of Poiseuille flow, shown in Fig. 1(b). This setup has been used successfully to predict the shear viscosity of supercritical $CO_2$ with polymers as viscosifying agents,[49] and also in studies of viscosification of solutions with branched copolymers.[34]

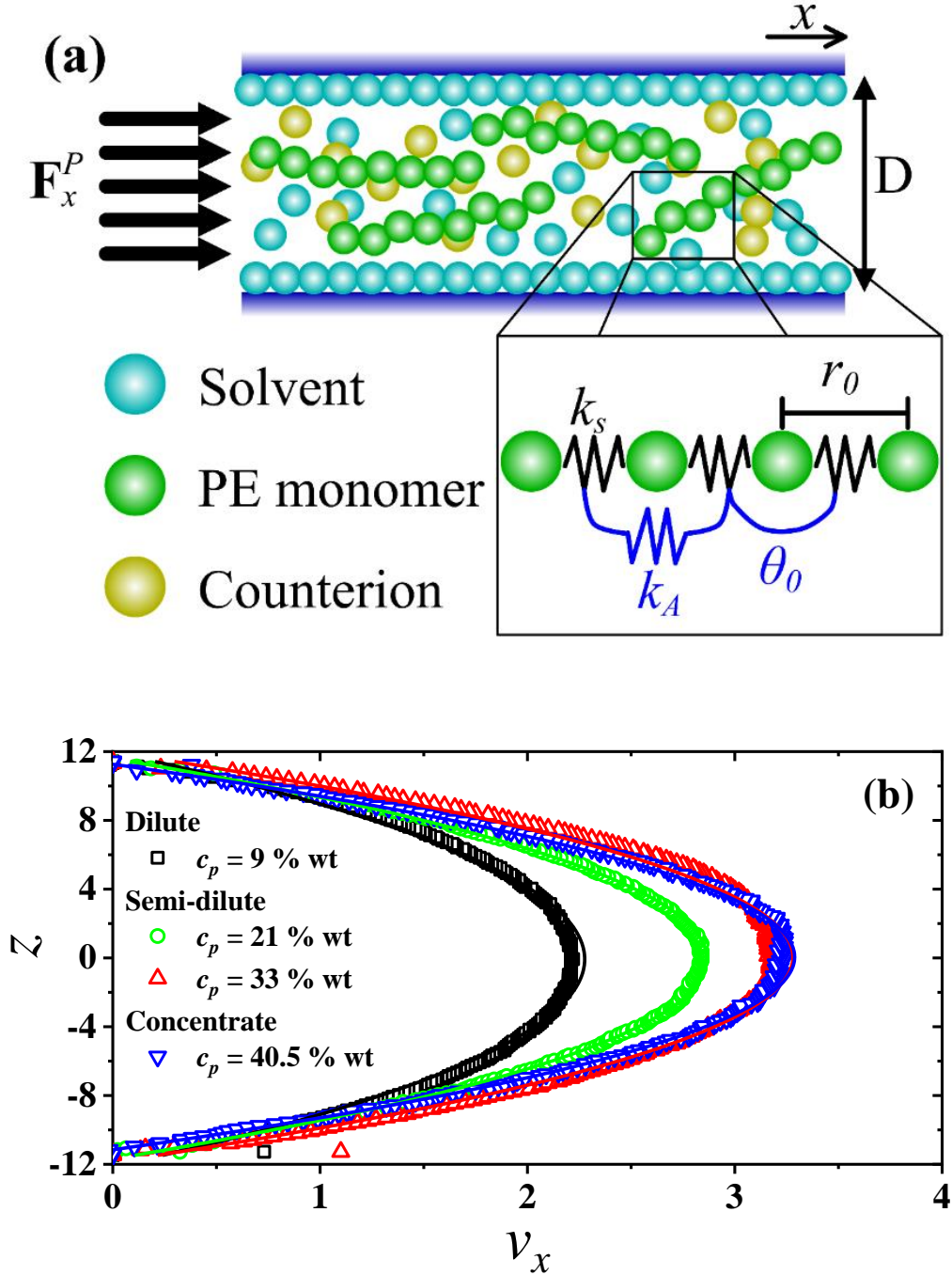

**Fig. 1**. (a) Illustration of the setup implemented in this work for confined PEs under a Poiseuille flow. A thin layer of solvent particles is formed close to the walls ($z \leq 0.15\, r_c$), having a velocity equal to zero along the $z$-direction. The linear and angular bonding in PEs is depicted as well, wherein the constants $k_s = 100$, $r_0 = 0.7 r_c$ (see eq. (1)) and $\theta_0 = 180°$

(see eq. (2)) are fixed in all cases, varying only the values of $k_A$ (see Table 2). (b) Profiles of the *x*-component of the velocity of solvent beads under stationary flow, along the *z*-direction of the simulation box, perpendicular to the direction of the external force ($\mathbf{F}_x^P$) producing the flow for salt-free systems for different polyelectrolyte concentration ($c_p$, in wt %) and $L_p = 40.47\ nm$. Solid lines are the parabolic fits (see eq. (5)). The axes are expressed in reduced units.

The viscosity ($\eta$) of the system is obtained directly after fitting the analytic velocity profiles with the steady-state solution of the Navier-Stokes equation:[38]

$$v_x(z) = \frac{\rho F_x^P}{2\eta}(z - z_0)(D - z_0 - z) \tag{5}$$

where, $\rho$ is the reduced number density of the fluid, $F_x^P$ is the magnitude of the external force applied on the DPD particles to produce the flow, $D$ is the separation distance between walls, and $z_0$ is the position at which the extrapolation of the velocity profile reaches 0 value. Fitting the velocity profiles obtained from the simulations to eq. (5), $\eta$ can be readily obtained, as seen in Fig. 1(b).

The electrostatic interaction is modeled by means of the standard Ewald sums method,[50,51] adapted for DPD method using charge distributions instead of point charges.[52] The charge distribution used here is a Slater-type, radially decaying exponential with decay length λ, and given by[53]

$$\rho(r) = \frac{q}{\pi \lambda^3} e^{-2r/\lambda}. \tag{6}$$

The magnitude of the reduced electrostatic force between two charge distributions with valence $Z_i$ and $Z_j$ separated by a distance $r^* = r_{ij}/r_c$ is given by

$$F_{ij}^{E^*}(r^*) = \frac{\Gamma Z_i Z_j}{4\pi r^{*2}}\{1 - [1 + 2\beta^* r^*(1 + \beta^* r^*)]\exp(-2\beta^* r^*)\} \tag{7}$$

where, $\Gamma = e^2/(k_B T \varepsilon_0 \varepsilon_T r_c)$, $\beta^* = r_c \beta = 5 r_c/(8\lambda)$; $e$ is the electron charge, $k_B$ is Boltzmann' constant, $\varepsilon_0$ is vacuum's permittivity, and $\varepsilon_T = 78.3$ is the water relative dielectric permittivity at room temperature. This charge distribution model avoids the formation of artificial ionic pairs, which would lead to a singularity in the Coulomb interaction when their relative distance becomes zero. It has been used successfully in DPD simulations to model PEs brushes,[53] the self-assembly of PEs in aqueous media,[54] and adsorption processes of confined PEs.[55]

## III. SIMULATION DETAILS

All simulations were carried out under canonical ensemble conditions, i.e., at constant number density and temperature, where the reduced number density is $\rho = 3$, and the temperature is always fixed at $k_B T = 1$, which is also the reduced unit of energy. The intensity of the random force is $\sigma = 3$, and dissipation parameter is $\gamma = 4.5$.[36] The cutoff radius is the reduced unit of length, and it is $r_c = 1 \approx 6.46$ Å for a coarse-grained degree of three. The PEs were modeled as linear chains made up of 15 DPD particles each, freely joined by harmonic springs, varying the PE concentration from $c_P = 1$ wt % to $c_P = 42$ wt %. The dimensions of the simulation box were set in all cases as $l_x = l_y = 12 \, r_c$ and $l_z = 23 \, r_c$, with the total number of DPD particles equal to $N = 10^4$.

The number of counterions in the system is the same as the number of PE monomers, and four different values for the coarse-grained salt ions charges were chosen to change the ionic strength, $Z = 0, 6, 9$ and $12$, all in units of the elementary charge, $e$. The parameters of the Ewald sums were $\alpha = 0.15$ Å$^{-1}$, the maximum vector $k^{max} = (5,5,5)$, $\beta = 0.929$ and $\lambda = 6.95$ Å.[51] The solvent quality was determined by the parameters of the conservative DPD force, $a_{ij}$, between solvent and polyelectrolyte monomer beads, choosing $a_{ij} = 60$ to model good solvent, and $a_{ij} = 78.3$ for theta solvent conditions, see Table 1. All simulations were run for 15 blocks of $10^4$ time steps each, using a time step of $\Delta t = 0.03 \approx 0.1 ps$. The first 5 blocks of each simulation are used to reach stationary flow, while the last 10 blocks were used to the production phase, where properties are averaged.

## IV. RESULTS AND DISCUSSION

The effect of the flexibility of the PE chains in the performance of the viscosity is studied, under good and theta solvent conditions, with and without salt as a function of the PE concentration under Poiseuille flow; see setup in Fig. 1 for details. The results are discussed in the next sections. The results are presented in two sections, one for each solvent quality and for each case, with and without salt. The effect of increasing the persistent length ($L_p$) at different concentration ($c_p$ in wt %) is studied, to explore the different scaling regimes.

**(a) Good solvent salt free system**

Figure 2 shows the results of the viscosity of different $L_p$ values (see eq. (3)) as a function of PE concentration in salt-free, good solvent conditions, using parameters presented in Table 1. The relative viscosity ($\eta/\eta_0$), defined as the ratio of the viscosity of PE in solution over the viscosity of the pure solvent, $\eta_0$, grows as a function of PE concentration for: flexible ($L_p = 0.93, 1.58$ and $4.29\ nm$, black, red and blue curves in Fig. 2, respectively) and rigid ($L_p = 90.21$ and $94.74\ nm$, green and olive curves in Fig. 2, respectively) PE structures. However, non-monotonous behavior is observed for semiflexible structures ($L_p = 40.47, 44.97, 49.52$ and $67.60\ nm$, navy, violet, purple and wine curves in Fig. 2, respectively), where an increase in the viscosity is initially observed in a diluted regime ($c_p <$ 20 $wt$ %) and in the concentrated regime ($c_p > 35\ wt$ %), followed by a decrease in the viscosity for the semi-diluted (20 $wt$ % $< c_p < 35$ % $wt$), see Fig. 2.

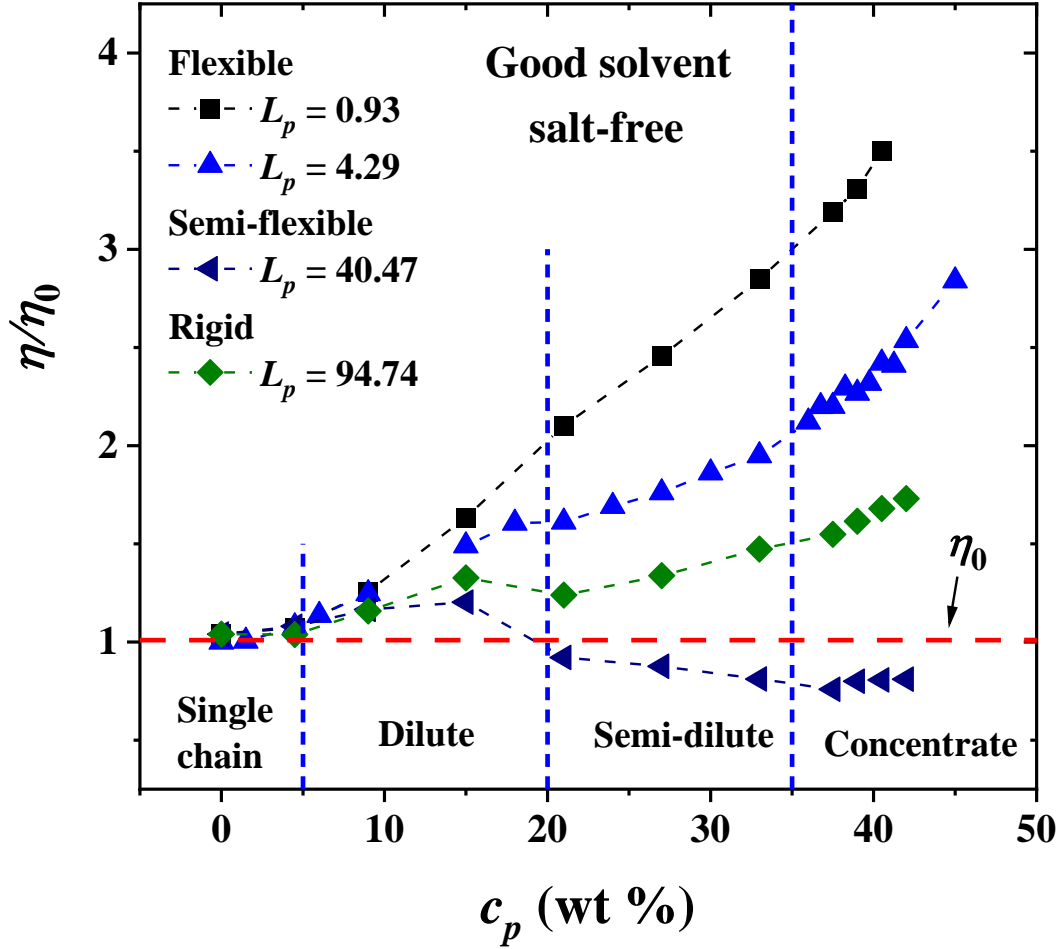

**Fig. 2**. Relative viscosity $\eta/\eta_0$ as a function of polyelectrolyte concentration, $c_p$, under salt-free, good solvent conditions. The values of the persistence length $L_p$ are reported in *nm*.

To understand this behavior, Fig. 3 presents snapshots and density profiles for the phases obtained in our systems depending on the flexibility of PEs in the semi-diluted region, at fixed PE concentration ($c_p = 33\ wt$ %). In the dilute regime the chains do not overlap, they are far from each other and at very low concentrations can be considered as a single chain problem. For this reason, the effect of the flexibility is negligible and the behavior of the relative viscosity as a function of concentration is almost the same. For semi-dilute and concentrated solutions, the PEs start to overlap, and their conformation is clearly modified. This case is especially critical in practice because it is related to the condensation threshold emerging different conformations: hexagonal (canonical) and isotropic[19] and the change in the solution's viscosity is very different from the dilute regime. In the absence of salt, the electrostatic repulsions between monomers are long ranged in the semi-dilute regime and the entropy contributions due to chain flexibility are not very significant. The semi-dilute solutions can form parallel stackings of rods, creating a phase which is macroscopically uniaxial as seen for flexible and semi-flexible PEs in Fig. 3(b) and 3(c), closer to the hexagonal phase also known as canonical phase.[19] This type of configuration is found for semiflexible PEs ($22.38 \leq L_p \leq 67.60\ nm$), see Fig. 3(c). The other possible configuration is the so-called random isotropic phase (see Fig. 3(a) and 3(d)), consisting of overlapping, partially flexible PEs, which can be analyzed by scaling methods.[19]

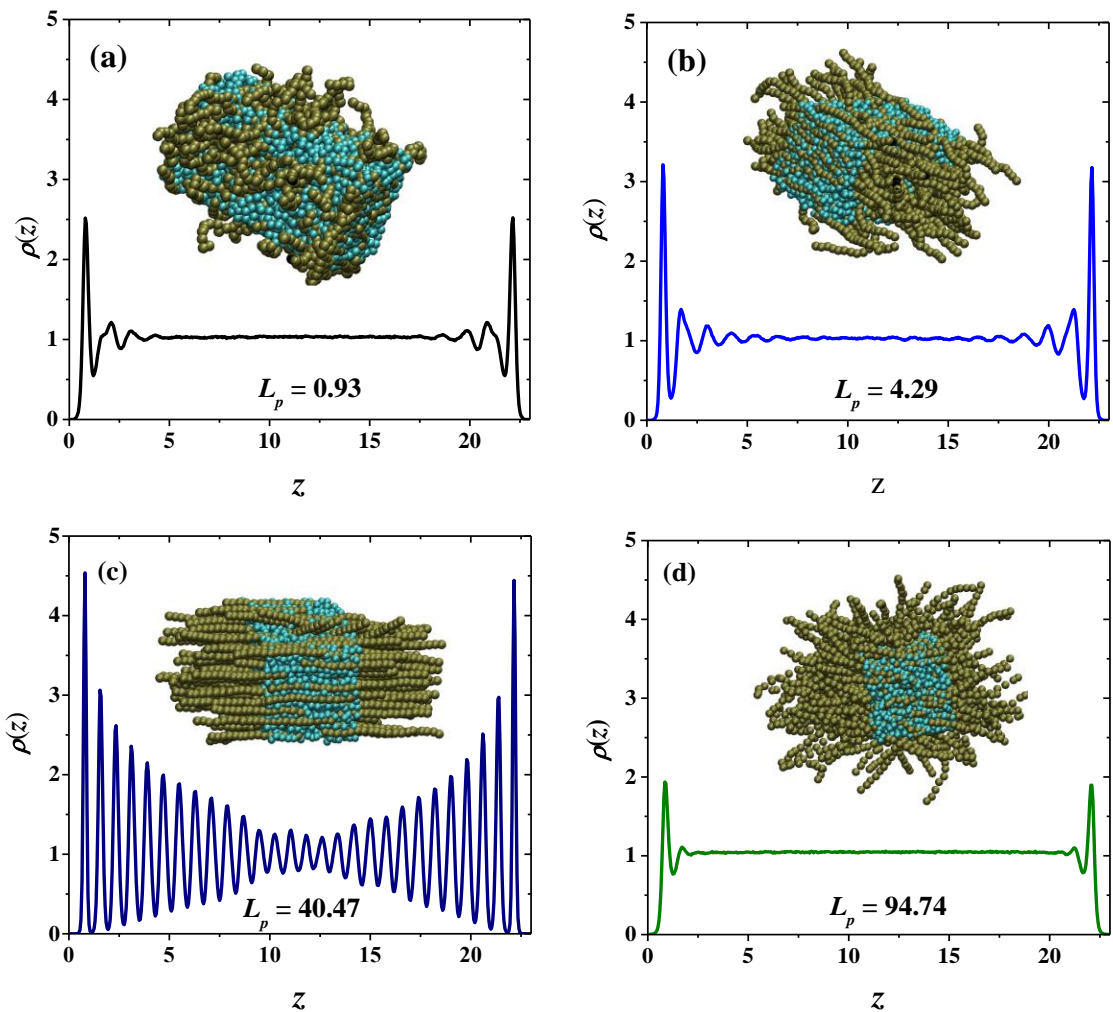

**Fig. 3.** Density profiles of the PEs and snapshots of systems with (a) $L_p = 0.93\ nm$, (b) $L_p = 4.29\ nm$, (c) $L_p = 40.47\ nm$ and (d) $L_p = 94.74\ nm$. The PE concentration in all cases is $c_P = 33$ wt %. In the snapshots the PEs are shown in dark green and the solvent beads in cyan.

Fig. 4 presents the radial distribution functions (g(r)'s) between the solvent ($CO_2$) and the monomers of the PEs for three characteristic persistence lengths belonging to the three different stiffness regimes: the flexible regime ($L_p = 4.29\ nm$, Fig. 4(a)), semi-flexible regime ($L_p = 40.47\ nm$, Fig. 4(b)) and rigid regime ($L_p = 94.74\ nm$, Fig. 4(c)). The radial distribution functions for three different PE concentrations are shown (dilute, semi-dilute and concentrated regime, $c_p = 9, 33$ and $40.5$ wt %, respectively) to investigate the effects on the spatial distribution between $CO_2$ and the PEs produced by the PE stiffness. Without salt, the spatial ordering between $CO_2$ and PEs is reduced when the concentration of PEs increase. This effect is notorious when the PEs exhibit an isotropic phase for rigid polymers (see Fig. 3(d) and Fig. 6 (a)), because the $CO_2$ fails to form layers around the PEs due to the entanglement between the PEs. Therefore, the PEs at the concentrated regime tend to thicken the $CO_2$. On the other hand, and contrary to what happen in the rigid regime, in the semi-flexible regime (Fig. 4(b)) the $CO_2$ and PEs become structured, as expected for the hexagonal phase (see Fig. 3(c)). Such structure between the $CO_2$ and PEs is responsible for the PEs working as $CO_2$ thinners at this stiffness of the PEs. These results support the fact that in the dilute regime the effect of the PE stiffness on the viscosity is negligible since the spatial distribution between $CO_2$ and PEs are similar. This feature can be seen from the g(r) functions of systems with $c_p = 9$ wt % (black curves in Fig. 4).

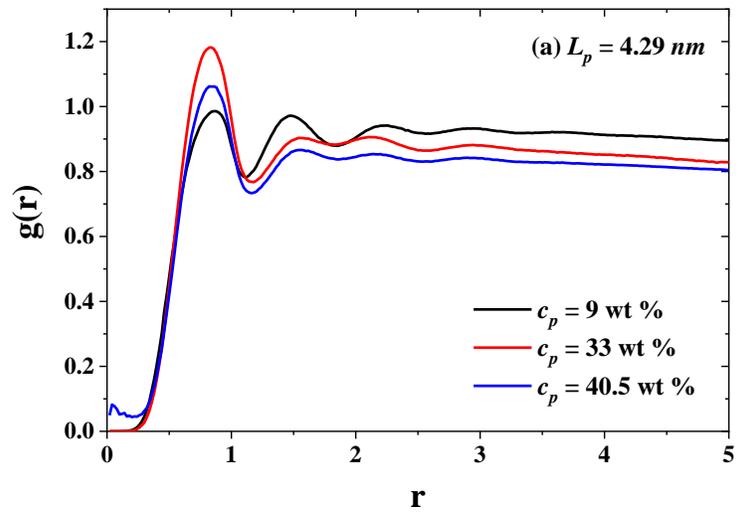

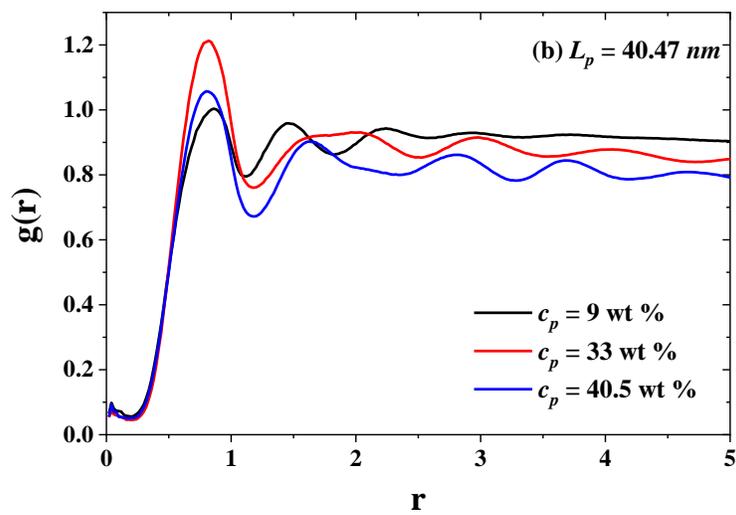

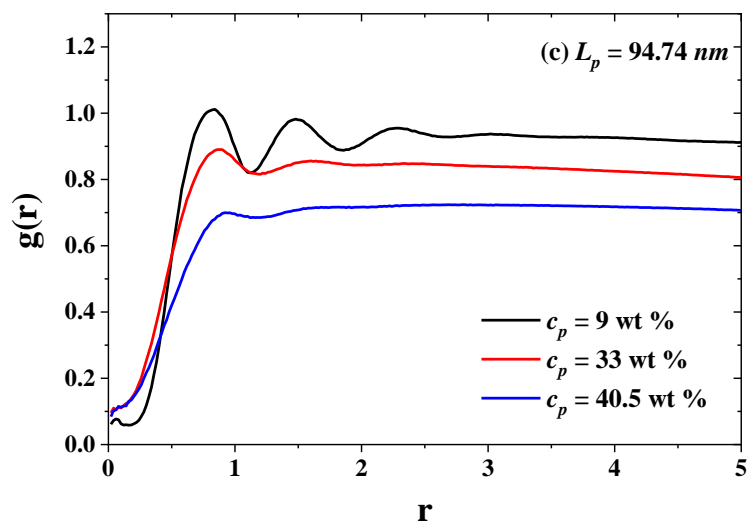

**Fig. 4**. Radial distribution functions between the solvent ($CO_2$) and the polyelectrolytes (PEs) for three persistence lengths values: (a) flexible PEs, (b) semi-flexible PEs, and (c) rigid PEs, at three different concentrations belonging to the dilute (black curves), semi-dilute (red curves), and concentrated (blue curves) regimes.

**(b) Scaling in the Isotropic phase: Flexible and rigid polyelectrolytes**

The viscosity in the isotropic phase obeys scaling laws. Figure 5 shows the logarithmic plot of the dependence of the relative viscosity with PE concentration, for the semi-dilute and concentrated regimes, and the exponents (*s*) of the scaling $\eta/\eta_0 \sim c_P^s$. For the semirigid and rigid polyelectrolytes with $L_p = 4.29 \, nm$ and $94.74 \, nm$, respectively, the classical scaling exponent for the viscosity $\eta$ of semi dilute, entangled PE solutions from the empirical Fuoss law is obtained $\eta \sim c_P^{1/2}$ in the salt-free regime, see Fig. 5(a).[23] For more flexible PEs, where the entropic contribution becomes important ($L_p = 0.93 \, nm$ and $1.58 \, nm$), the exponent $s = 0.68$ is found; see the circles and squares in Fig. 5(a). This prediction is in agreement with the experimental results for sodium carboxymethylcellulose (NaCMC)[20–22] and acrylamide-sodium-2-acrylamido-2-methylpropane-sulfonate (AM-NaAMPS).[23]

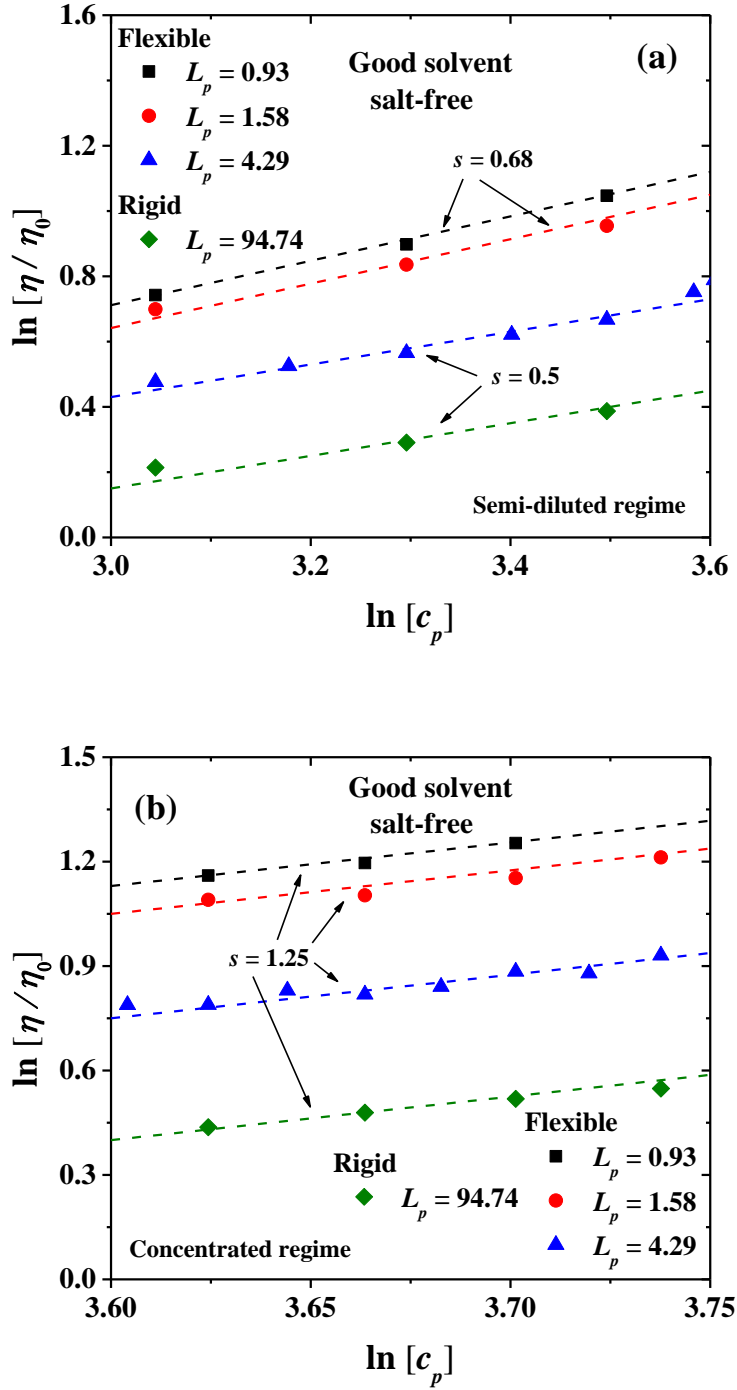

**Fig. 5**. Scaling of the relative viscosity of the solvent, $\eta/\eta_0 \sim c_P^s$, with PE concentration for the semi-diluted (a) and concentrated regimes (b), along with their scaling exponents. The symbols are data from our numerical simulations and the dashed lines are best fits to the scaling law.

In the concentrated regime all the PE systems show the scaling exponent $s = 5/4 = 1.25$, in agreement with the expected behavior from Fuoss law for PE; see Fig. 5(b).[2,3] The scaling model of a PE chain in semi-diluted solutions is based on the hypothesis of the existence of a single length scale, namely the correlation length $\xi$. Theoretical and experimental studies[2,14,20] find that the correlation length depends on polymer concentration as $\xi \sim c_P^{-1/2}$. On length scales up to the correlation length $\xi$, the conformations of a PE chain in semi diluted solutions are presumed to be Gaussian, while for length scales larger than $\xi$ the PE chain is expected to be flexible and scaling properties emerge.[2,14] The hypothesis of a single length scale in semi dilute PE solutions is supported by experiments and molecular dynamic simulations[2,14] showing that $L_p$ and $\xi$ are proportional. Figure 6 shows the dependence of the relative viscosity on the persistent length, which offers additional information with respect to its dependence on concentration (Fig. 2). Three regimes are found, associated with flexible, semi-flexible, and rigid PE chains. The relative viscosity in semi-diluted and concentrated regimes shows a power law decay for small persistence lengths ($L_p < 20\ nm$), where the PEs are flexible. For semi-flexible PEs ($20\ nm < L_p < 67.6\ nm$) the viscosity remains nearly $L_p$ independent, while for rigid PEs ($L_p > 67.6\ nm$) the viscosity increases with $L_p$; see Fig. 6(a). The data displayed in Fig. 6(b) show that the relative viscosity in the semi dilute and concentrated regime does scale as $\eta/\eta_0 \sim L_P^{-1/2}$, as expected.[2,14]

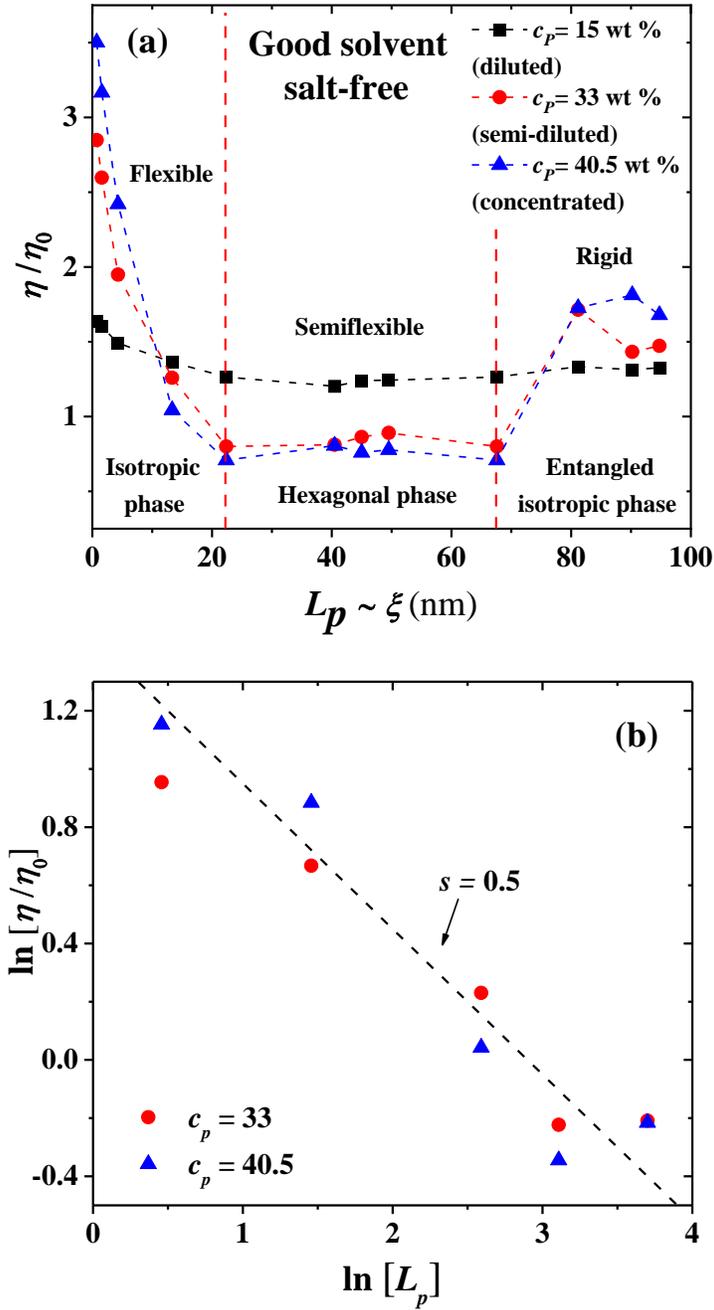

**Fig. 6**. (a) Influence of the persistence length $L_p$ (in nm) on the relative viscosity $\eta/\eta_0$, in salt-free, good solvent conditions, for three PE concentrations in three regimes, as labeled in the legend. (b) Scaling of the relative viscosity with $L_p$ for two PE concentrations, one in the semi-dilute ($c_P = 33$ wt %, blue squares), and one in the concentrated ($c_P = 40.5$ wt %, red circles) regimes. Red squares and blue circles represent the data from our simulations; the dashed line is the best fit to the scaling law $\eta/\eta_0 \sim c_P^s$, with $s = 0.5$.

**(c) Good solvent with salt**

Next, the behavior of PEs with salt added to the solution is explored. Figure 7 displays the results of the viscosity for four $L_p$ values as the PE concentration is increased. It is found that the relative viscosity increases as a function of the PE concentration for flexible ($L_p = 0.93\ nm, 4.29\ nm$) and rigid ($L_p = 90.21\ nm$) chains, as in the salt-free case (see Fig. 2). For semi-flexible PEs, such as those with $L_p = 44.99\ nm$, an increase in the viscosity is found in the single chain and diluted regime ($c_p < 20$ wt %) and a decrease in the viscosity for semi diluted and concentrated regime ($c_p > 20$ wt %). This is due to the different spatial conformations acquired by these systems; see the discussion of Fig. 2. At the largest PE concentration, the relative viscosity for $L_P = 0.93\ nm$ is even larger than in the solvent-free case, Fig. 2. The salt ions screen the Coulomb interaction and the PEs behave as flexible neutral polymers.

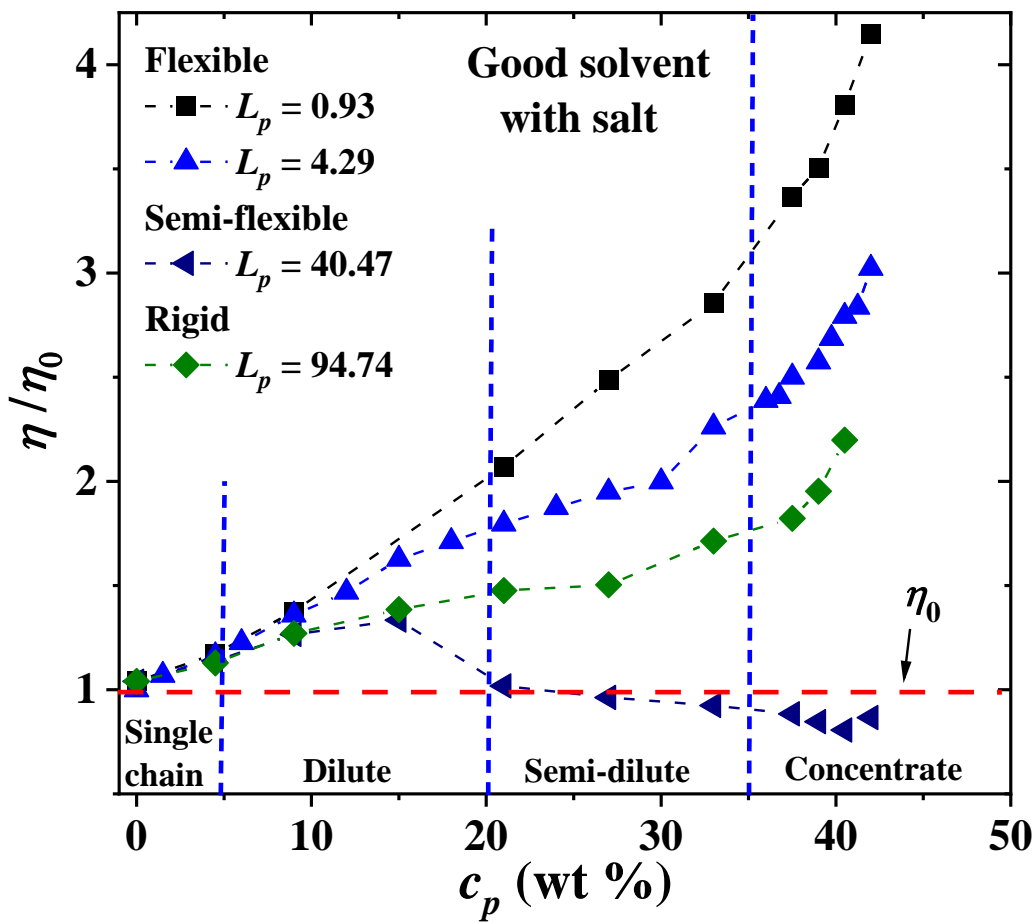

**Fig. 7**. Dependence of the relative viscosity on PE concentration, as the persistence length is increased, with salt ions of net charge of $Z = 6$ added to the solution. The viscosity of the pure solvent, without PEs, is $\eta_0$. The persistence length is reported in nm.

In Fig. 8, the radial distribution functions (g(r)'s) between the $CO_2$ solvent and PEs of three different stiffness regimes are presented: flexible regime, semi-flexible regime and rigid regime. Additionally, the g(r)'s for the dilute, semi-dilute and concentrated regimes are presented in Fig. 8 as well. The spatial ordering between $CO_2$ and PEs is reduced when the concentration of PEs increase, especially when the PEs is in the isotropic rigid phase where the $CO_2$ viscosity decreases. In the semi-flexible (Fig. 8(b)) regime the $CO_2$ and PEs produce a structured hexagonal phase and the PEs work as $CO_2$ thinners at this stiffness regime of the PEs.

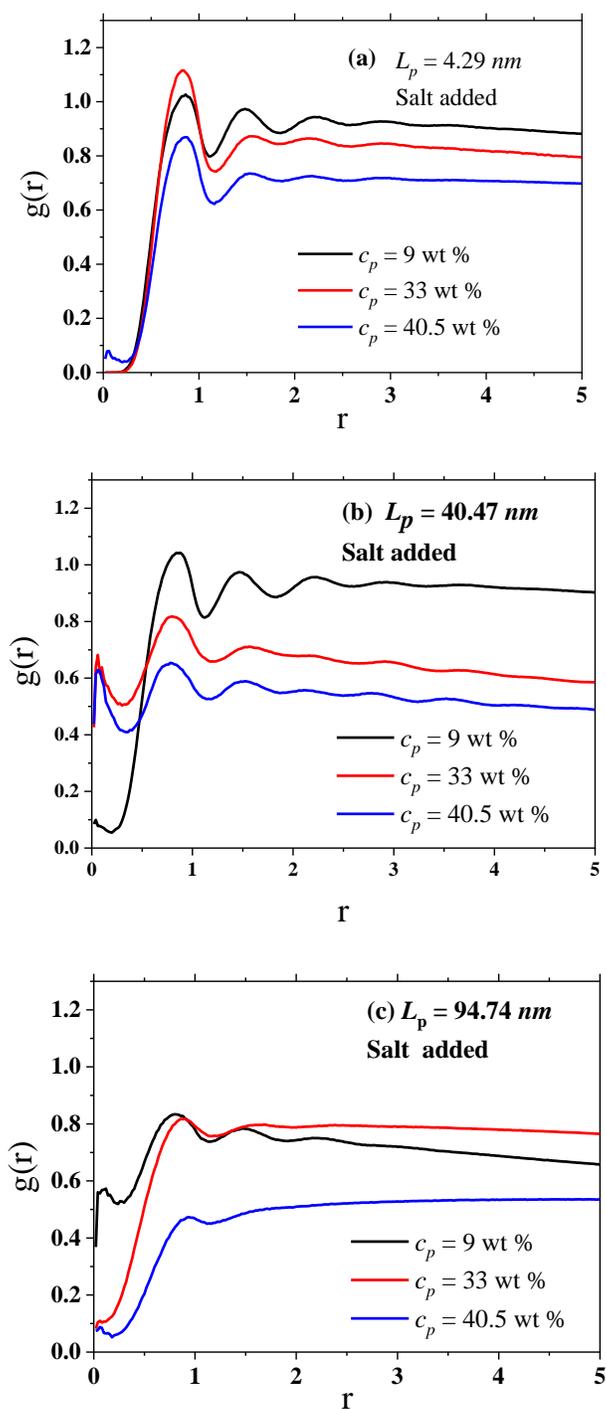

**Fig. 8**. Radial distribution functions between the solvent ($CO_2$) and the polyelectrolytes (PEs) for three persistence lengths values: (a) flexible PEs, (b) semi-flexible PEs, and (c) rigid PEs,

at three different concentrations belonging to the dilute (black curves), semi-dilute (red curves), and concentrated (blue curves) regimes.

For semi-dilute and concentrated solutions, the scaling exponents with salt added in the isotropic phase for flexible and rigid PEs are presented in Fig. 9. The exponents obtained follow Fuoss law $\eta \sim c_p^{3/2}$ when salt is added,[2,3] as has often been reported for sodium polystyrene sulfonate (NaPSS) solutions.[31,56–58]

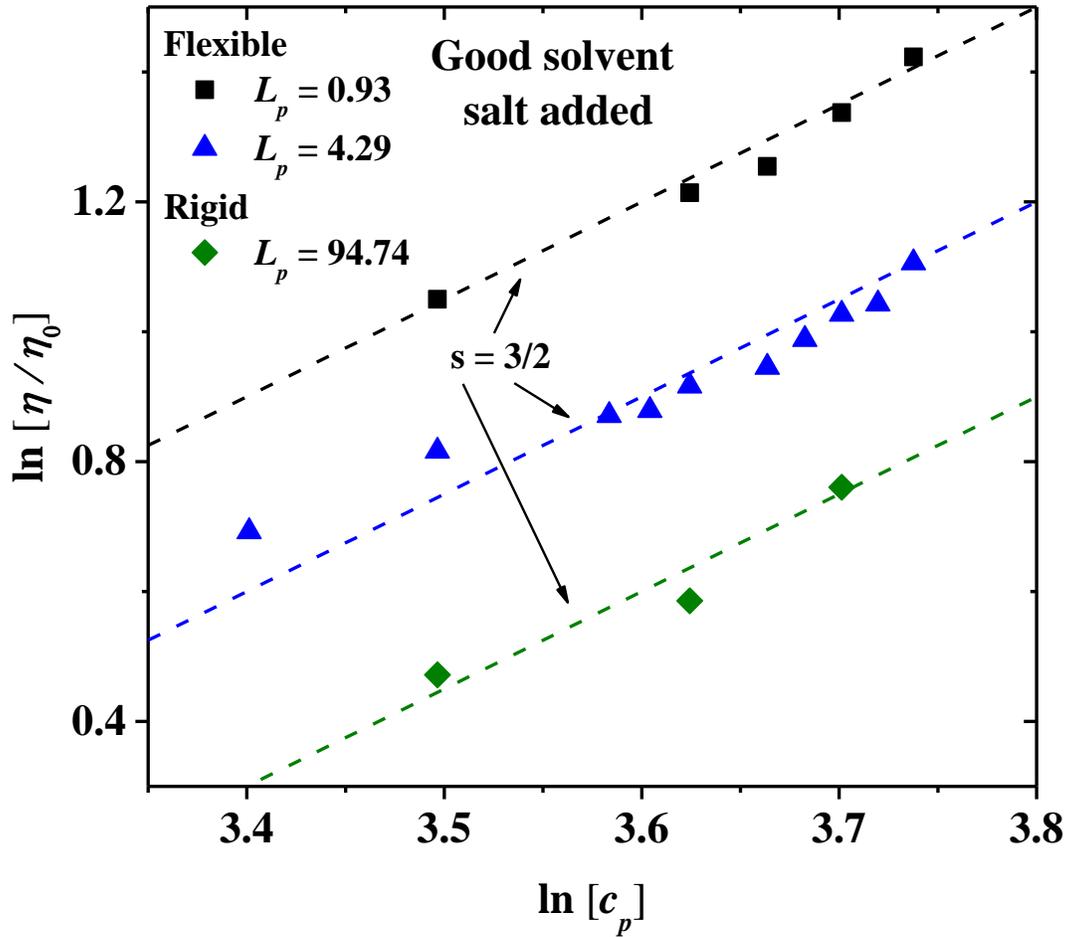

**Fig. 9**. The relative viscosity for PEs in solution with salt ions with charge of $Z = 6$ under good solvent conditions. The scaling exponent, $s = 3/2$, follows Fuoss law $\eta \sim c_p^{3/2}$, for the concentrated regime ($c_P \geq 35$ wt %) with salt added, for three values of the persistence length.

Next, we explore in Fig. 10 the effects of increasing the charge of salt ions, for solutions with flexible PEs ($L_p = 4.29\ nm$) in the isotropic phase. To do this, the ionic concentration is kept constant while the charge of the ions is increased. A general increasing behavior in the viscosity is obtained when the charge of the ions increases, with respect to the ion-free solutions under good solvent conditions (Fig. 2). The three regimes (dilute, semi-dilute and concentrated) can be distinguished in the relative viscosity as indicated in Fig. 10. In the diluted region, the viscosity is a linear function of $c_p$, while for the semi-diluted and concentrated regimes a power law dependence emerges. The scaling behavior is more clearly borne out in Fig. 11.

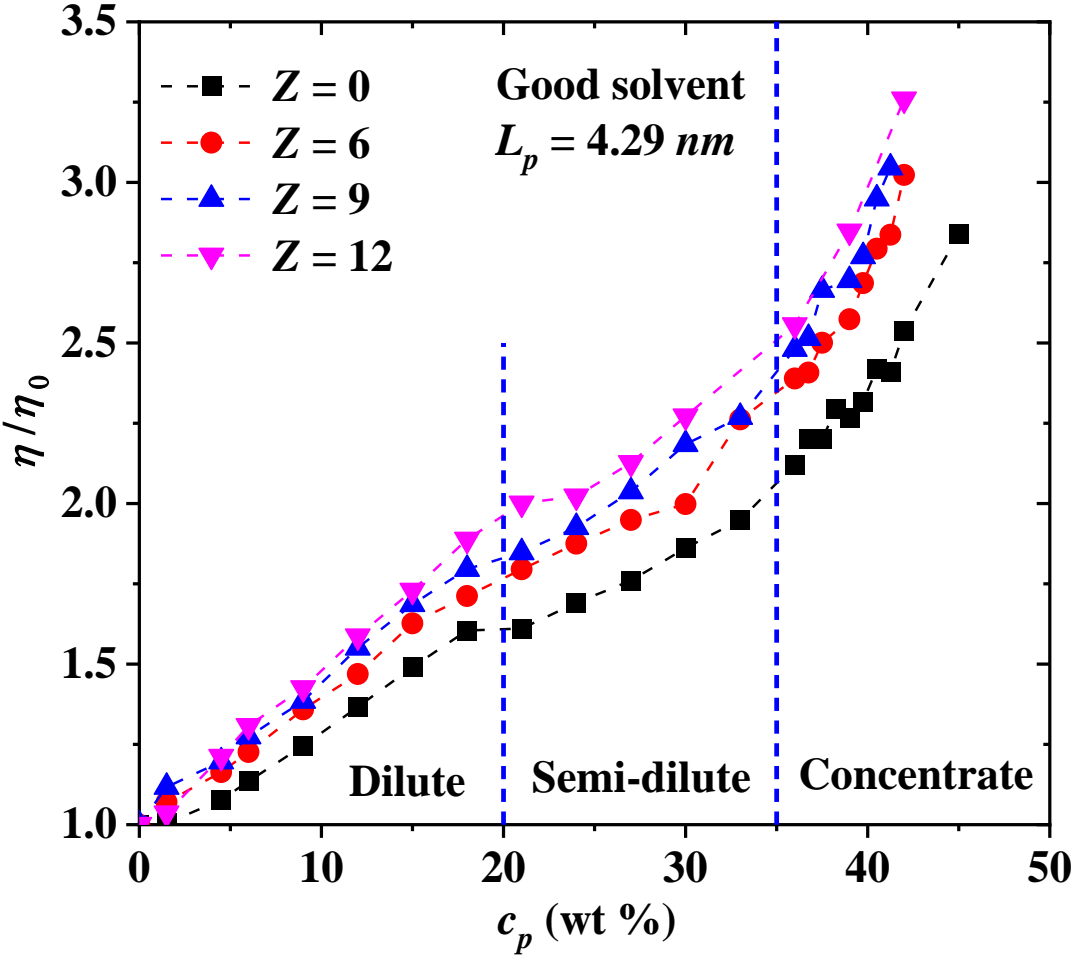

**Fig. 10.** Influence of ion's charge on the relative viscosity of PEs in solution, as a function of PE concentration. In all cases the persistence length is fixed at $L_P = 4.29\ nm$. The system is under good solvent conditions.

Figures 11(a) and 11(b) show the scaling exponents obtained for the relative viscosity as a function of $c_p$ in the semi-diluted and concentrated regimes, under different ion's charges. In the semi-dilute regime, the classical scaling exponent $s = 1/2$ for the viscosity is found, once again in agreement with Fuoss law $\eta \sim c_p^{1/2}$, with and without salt; see also Fig. 5(a). However, in the concentrated regime with increasing ion charge the viscosity has a stronger than linear concentration dependence, $\eta \sim c^{3/2}$. This scaling exponent differs from the case without salt where we obtain the $\eta \sim c^{5/4}$, which is the same scaling as that of the semi diluted unentangled neutral polymers in good solvent[52], as discussed in the previous section.

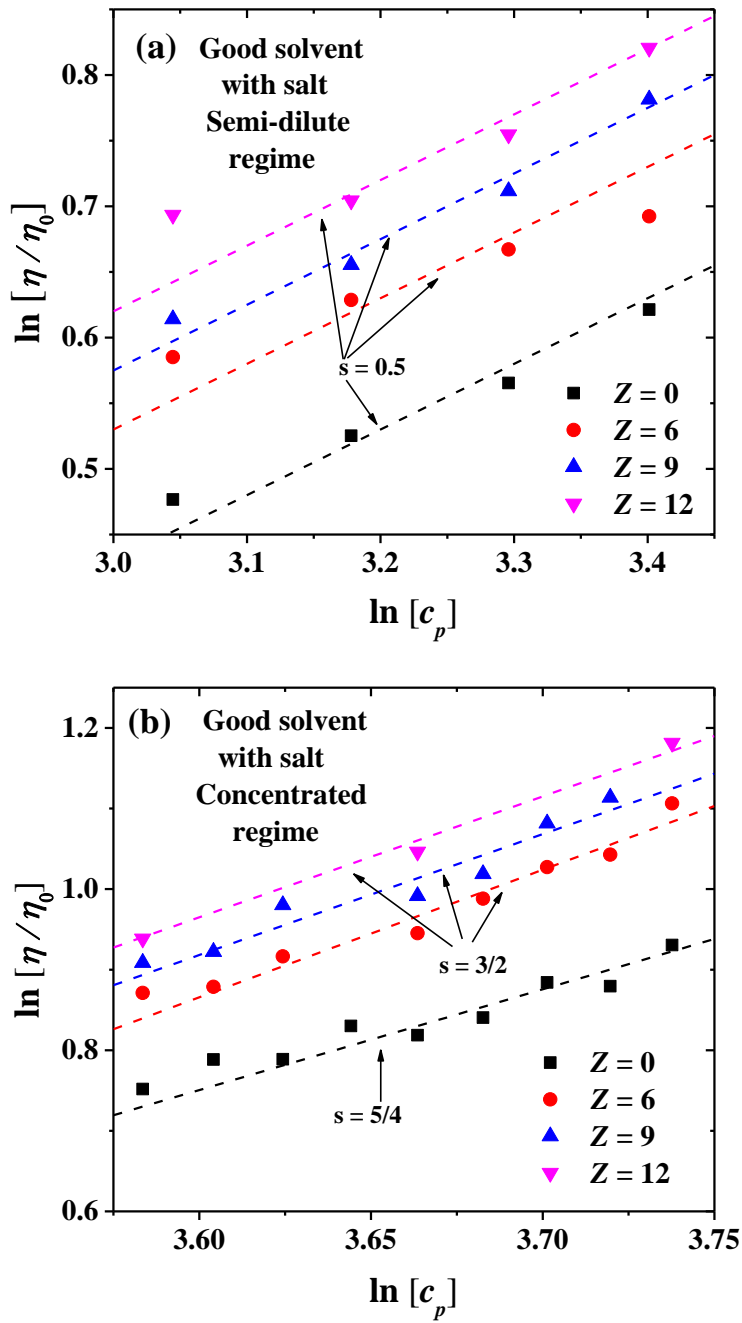

**Fig. 11.** Logarithmic plot of the relative viscosity as a function of polyelectrolyte concentration at different ionic strength. (a) Scaling exponents for semi-dilute regime, (b) scaling exponents for concentrated regime.

**(d) Salt-free system under theta solvent conditions**

Figure 12(a) shows the results of the viscosity at different $L_p$, as a function of polyelectrolyte concentration in theta solvent. The relative viscosity is found to grow as a function of the polymer concentration for flexible ($L_p = 0.93$ and $4.29\ nm$) and rigid ($L_p = 225.87\ nm$) polyelectrolytes structures. For semi-flexible structures ($44.99\ nm \leq L_p \leq 94.74\ nm$), non-monotonic behavior is obtained, where the PEs tend to thicken the $CO_2$, by increasing the relative viscosity. However, as the concentration is increased their relative viscosity decays to lower values, in particular for solutions with PEs with $L_p = 44.99\ nm$, to lower values than the intrinsic viscosity of $CO_2$ ($\eta_0$).

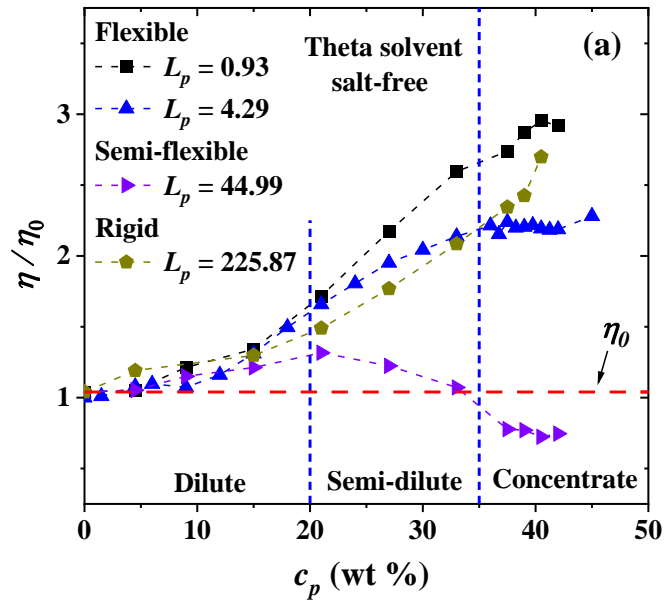

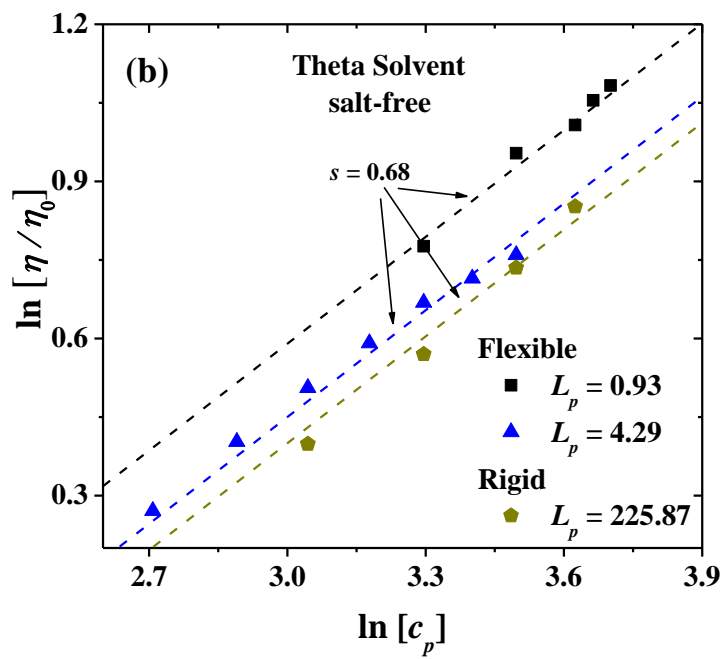

**Fig. 12**. (a) Relative viscosity of salt-free PEs in solution with different $L_p$ values as a function of polyelectrolyte concentration under theta solvent conditions. (b) Logarithmic dependence of the relative viscosity with the polyelectrolyte concentration in theta solvent. The expected scaling exponent is $s = 0.68$. In both panels the persistence length is reported in nm.

Figure 12(b) shows the logarithmic dependence of the viscosity with the polyelectrolyte concentration from the dilute to the concentrated regime in theta solvent, where scaling is expected to occur. For the flexible and rigid polyelectrolyte in the salt-free regime the expected scaling exponent is $s = 0.68$, similar to the one found for the very flexible polyelectrolytes in good solvent discussed in the previous section. In this case, the exponent for Fuoss law is not found, which is indicative that no entanglement between PEs take place under theta solvent conditions.

**(e) Theta solvent with salt added**

The behavior of the relative viscosity of theta solvent with PEs and salt added is presented in Fig. 13(a), when the PE´s concentration is increased. It is found that the relative viscosity increases as a function of the PE concentration for flexible and rigid chains, as in the salt-free case (see Fig. 12 (a)). For semi-flexible PEs, an increase in the viscosity is found in the single chain and diluted regimes but a decrease in the viscosity for semi diluted and concentrated regime is observed. Figure 13 (b) shows the scaling exponents obtained corresponding to Fuoss law.

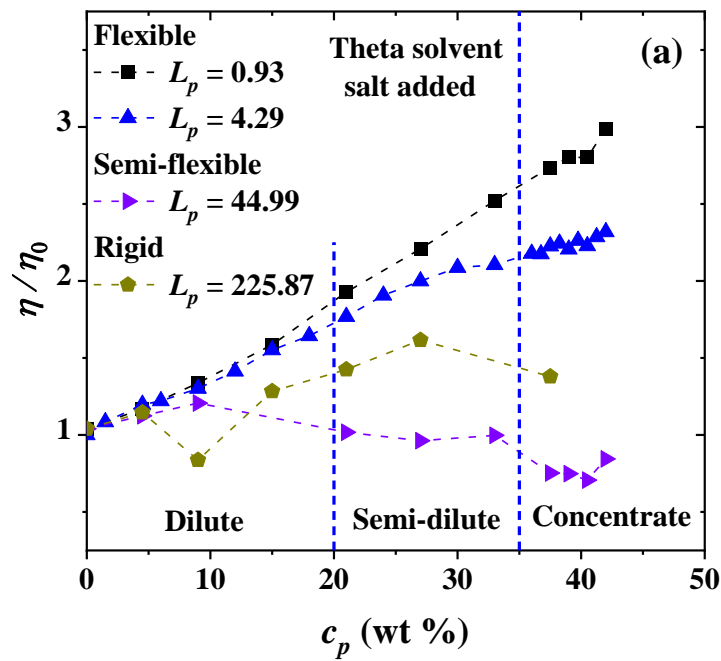

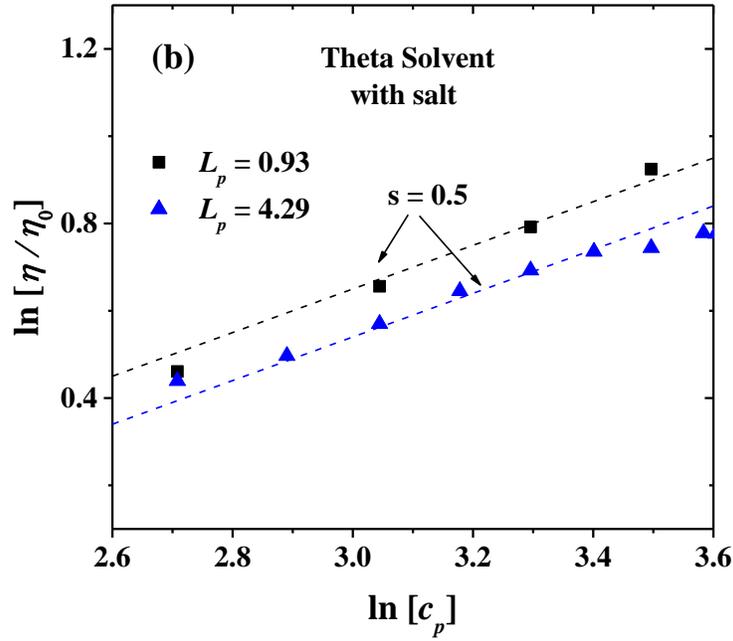

**Fig. 13.** (a) Relative viscosity of PEs with different $L_p$ values as a function of the concentration in theta solvent with salt ions added (charge of $Z = 6$), and (b) logarithmic dependence of the relative viscosity with the polyelectrolyte concentration in theta solvent with salt added. In both panels the persistence length is reported in nm.

Finally, we present the results for the effect of salt ions charge for flexible PEs with $L_p = 4.29\ nm$ in the isotropic phase in theta solvent (Fig. 14). The relative viscosity increases, in comparison with the salt-free system. However, the effect of the increase in the ions charge is almost the same as in the concentrated regime when salt is added, which appears to be a general trend under theta solvent conditions. The scaling exponents of the viscosity are shown in Fig. 14(b) as a function of $c_p$. When there is no salt in the semi-dilute regime, the scaling exponent is $s = 0.68$, and it change to $s = 0.5$ when salt is added, regardless the valence of the coarse-grained salt ions, reproducing the exponents for the Fouss law in good solvent.

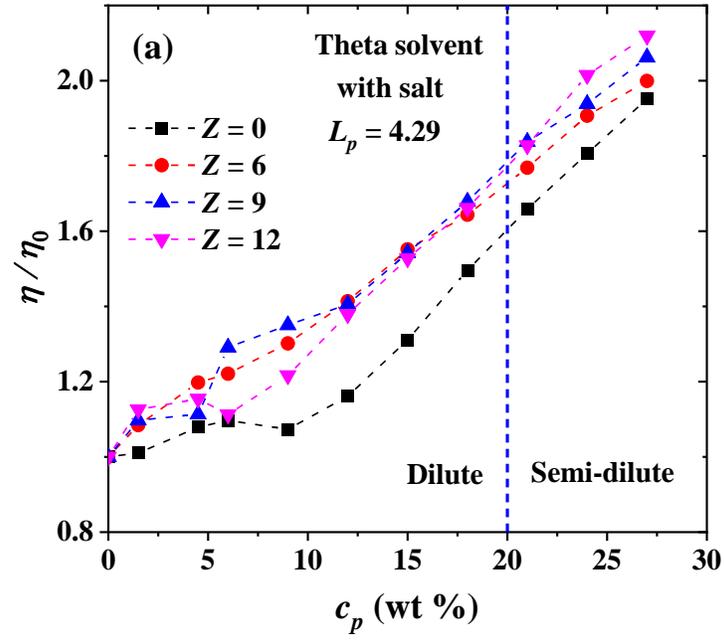

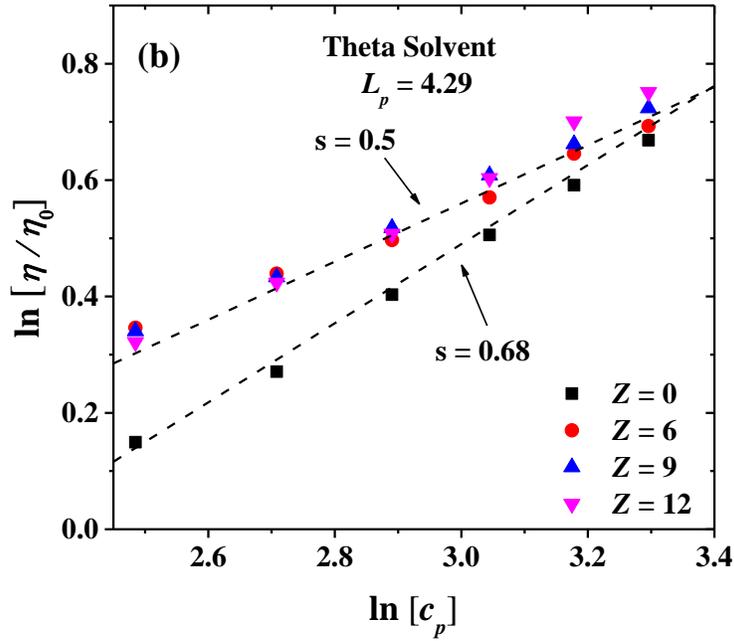

**Fig. 14.** (a) Logarithmic plot of relative viscosity as a function of polyelectrolyte concentration at different values of salt ions charges in theta solvent. (b) Scaling exponents of the relative viscosity for the semi-diluted regime. In both panels the persistence length is reported in nm.

## V. CONCLUSIONS

An extensive study of the scaling behavior on the viscosity of confined polyelectrolyte solutions under stationary Poiseuille flow, using electrostatic dissipative particle dynamics numerical simulations is presented here. The variables studied are polyelectrolyte chain flexibility (persistence length), concentration, solvent quality and ionic strength. The dynamics of PE solutions follows the same scaling hypothesis as neutral polymers do and it effectively relies on the existence of a single length scale, the correlation length $\xi$. Our analysis is based on the existence of this length scale, considering that it is the persistence length, $L_p$, which controls the properties of the solution via the flexibility of the polymer, which is essential for the complete understanding of the structure of the systems.

For salt-free polyelectrolyte chains in good solvent, our predictions agree with the scaling model of de Gennes, and with Rouse's model applied to semi-dilute solutions, following the empirical Fuoss law. The viscosity of salt-free unentangled, semi-dilute solutions is predicted to obey Fuoss law $\eta \sim c^{1/2}$. Above the entanglement concentration $c_e$ in salt-free solutions, we confirm the prediction for the viscosity of entangled solutions, $\eta \sim c_p^{3/2}$. This is in remarkable agreement with experiments performed in sulfonated polystyrene[59] and poly(ZV-methyl-2-vinylpyridinium chloride)[60] solutions, where the Fuoss law is confirmed by experiments. Additionally, the results for good and theta solvent conditions show that solvent quality is an essential factor influencing the chain's conformations due to the competition of electrostatic and non-electrostatic interactions between the PEs monomers and the solvent molecules.

Although there are some reports in the literature on the viscosity of PEs, very few are available on the influence of the PEs persistence length. This is essential to further the understanding of viscosity scaling in entangled polyelectrolyte solutions under different solvent conditions. This is needed to perform intelligent design of new polyelectrolytes, capable of tuning the viscosity in different complex fluids. Additionally, studying the effects of ionic strength using multivalent ions on complexation in PE solutions is of great importance because this phenomenon is present in biological systems. The strong salt effect on complexation confirms the electrostatic mechanism of viscosification in our simulations. Lastly, it is to be noted that the non-linear dynamics of polyelectrolyte solutions has been overlooked by theoretical approaches; for these reasons dynamic electrostatic coarse-grained computer simulations are promising tools to help in the design of new materials and in polymer processing.

## CONFLICTS OF INTEREST

There are no conflicts of interest to declare.

## ACKNOWLEDGMENTS

E. M. acknowledges funding from COMECYT, under the program FICDTEM-2021-013. The calculations reported here were mostly performed using the supercomputing facilities of ABACUS Laboratorio de Matemática Aplicada y Cómputo de Alto Rendimiento of CINVESTAV-IPN. The authors acknowledge also the computer resources offered by the Laboratorio de Supercómputo y Visualización en Paralelo (LSVP-Yoltla) of UAM-Iztapalapa, where part of the simulations was carried out. J. D. H. V. thanks CONACYT for a postdoctoral fellowship. A.G.G. thanks also CONACYT for support through grant 320197.

# REFERENCES

(1) Fuoss, R. M. Polyelectrolytes. *Discuss Faraday Soc* **1951**, *11*, 125. https://doi.org/10.1039/df9511100125.

(2) Dobrynin, A. v.; Colby, R. H.; Rubinstein, M. Scaling Theory of Polyelectrolyte Solutions. *Macromolecules* **1995**, *28* (6), 1859–1871. https://doi.org/10.1021/ma00110a021.

(3) Muthukumar, M. Dynamics of Polyelectrolyte Solutions. *Journal of Chemical Physics* **1997**, *107* (7), 2619–2635. https://doi.org/10.1063/1.474573.

(4) Cohen, J.; Priel, Z.; Rabin, Y. Viscosity of Dilute Polyelectrolyte Solutions. *J Chem Phys* **1988**, *88* (11), 7111–7116. https://doi.org/10.1063/1.454361.

(5) Rabin, Y.; Cohen, J.; Priel, Z. Viscosity of Polyelectrolyte Solutions—the Generalized Fuoss Law. *Journal of Polymer Science: Polymer Letters Edition* **1988**, *26* (9), 397–399. https://doi.org/10.1002/pol.1988.140260904.

(6) Cohen, J.; Priel, Z. Viscosity of Dilute Polyelectrolyte Solutions: Concentration Dependence on Sodium Chloride, Magnesium Sulfate and Lanthanum Nitrate. *Macromolecules* **1989**, *22* (5), 2356–2358. https://doi.org/10.1021/ma00195a060.

(7) Vink, H. Rheology of Dilute Polyelectrolyte Solutions. *Polymer (Guildf)* **1992**, *33* (17), 3711–3716. https://doi.org/10.1016/0032-3861(92)90660-O.

(8) Roure, I.; Rinaudo, M.; Milas, M.; Frollini, E. Viscometric Behaviour of Polyelectrolytes in the Presence of Low Salt Concentration. *Polymer (Guildf)* **1998**, *39* (22), 5441–5445. https://doi.org/10.1016/S0032-3861(97)10274-9.

(9) Carrington, S.; Odell, J.; Fisher, L.; Mitchell, J.; Hartley, L. Polyelectrolyte Behaviour of Dilute Xanthan Solutions: Salt Effects on Extensional Rheology. *Polymer (Guildf)* **1996**, *37* (13), 2871–2875. https://doi.org/10.1016/0032-3861(96)87653-1.

(10) Plucktaveesak, N.; Konop, A. J.; Colby, R. H. Viscosity of Polyelectrolyte Solutions with Oppositely Charged Surfactant. *J Phys Chem B* **2003**, *107* (32), 8166–8171. https://doi.org/10.1021/jp0275995.

(11) Konop, A. J.; Colby, R. H. Polyelectrolyte Charge Effects on Solution Viscosity of Poly(Acrylic Acid). *Macromolecules* **1999**, *32* (8), 2803–2805. https://doi.org/10.1021/ma9818174.

(12) Eckelt, J.; Knopf, A.; Wolf, B. A. Polyelectrolytes: Intrinsic Viscosities in the Absence and in the Presence of Salt. *Macromolecules* **2008**, *41* (3), 912–918. https://doi.org/10.1021/ma702054f.

(13) Dou, S.; Colby, R. H. Solution Rheology of a Strongly Charged Polyelectrolyte in Good Solvent. *Macromolecules* **2008**, *41* (17), 6505–6510. https://doi.org/10.1021/ma8001438.

(14) Rubinstein, M.; Colby, R. H.; Dobrynin, A. v. Dynamics of Semidilute Polyelectrolyte Solutions. *Phys Rev Lett* **1994**, *73* (20), 2776–2779. https://doi.org/10.1103/PhysRevLett.73.2776.

(15) Wyatt, N. B.; Liberatore, M. W. Rheology and Viscosity Scaling of the Polyelectrolyte Xanthan Gum. *J Appl Polym Sci* **2009**, *114* (6), 4076–4084. https://doi.org/10.1002/app.31093.

(16) DOBRYNIN, A.; RUBINSTEIN, M. Theory of Polyelectrolytes in Solutions and at Surfaces. *Prog Polym Sci* **2005**, *30* (11), 1049–1118. https://doi.org/10.1016/j.progpolymsci.2005.07.006.

(17) Colby, R. H. Structure and Linear Viscoelasticity of Flexible Polymer Solutions: Comparison of Polyelectrolyte and Neutral Polymer Solutions. *Rheol Acta* **2010**, *49* (5), 425–442. https://doi.org/10.1007/s00397-009-0413-5.

(18) Muthukumar, M. *50th Anniversary Perspective* : A Perspective on Polyelectrolyte Solutions. *Macromolecules* **2017**, *50* (24), 9528–9560. https://doi.org/10.1021/acs.macromol.7b01929.

(19) de Gennes, P.; Pincus, P.; Velasco, R.; Brochard, F.; de Gennes, P. G.; Velasco, R. M.; Bâtiment, F. Remarks on Polyelectrolyte Conforma-Tion. *Journal de Physique* **1976**, *37* (12), 1461–1473. https://doi.org/10.1051/jphys:0197600370120146100ï.

(20) Lopez, C. G.; Rogers, S. E.; Colby, R. H.; Graham, P.; Cabral, J. T. Structure of Sodium Carboxymethyl Cellulose Aqueous Solutions: A SANS and Rheology Study. *J Polym Sci B Polym Phys* **2015**, *53* (7), 492–501. https://doi.org/10.1002/polb.23657.

(21) Lopez, C. G.; Colby, R. H.; Graham, P.; Cabral, J. T. Viscosity and Scaling of Semiflexible Polyelectrolyte NaCMC in Aqueous Salt Solutions. *Macromolecules* **2017**, *50* (1), 332–338. https://doi.org/10.1021/acs.macromol.6b02261.

(22) Lopez, C. G. Entanglement Properties of Polyelectrolytes in Salt-Free and Excess-Salt Solutions. *ACS Macro Lett* **2019**, *8* (8), 979–983. https://doi.org/10.1021/acsmacrolett.9b00161.

(23) Kujawa, P.; Audibert-Hayet, A.; Selb, J.; Candau, F. Effect of Ionic Strength on the Rheological Properties of Multisticker Associative Polyelectrolytes. *Macromolecules* **2006**, *39* (1), 384–392. https://doi.org/10.1021/ma051312v.

(24) Dubrovskii, S. A.; Zelenetskii, A. N.; Uspenskii, S. A.; Khabarov, V. N. Effect of Borax Additives on the Rheological Properties of Sodium Hyaluronate Aqueous Solutions. *Polymer Science - Series A* **2014**, *56* (2), 205–210. https://doi.org/10.1134/S0965545X14020047.

(25) del Giudice, F.; Calcagno, V.; Esposito Taliento, V.; Greco, F.; Netti, P. A.; Maffettone, P. L. Relaxation Time of Polyelectrolyte Solutions: When µ -Rheometry Steps in Charge . *J Rheol (N Y N Y)* **2017**, *61* (1), 13–21. https://doi.org/10.1122/1.4965930.

(26) Hara, M. *Polyelectrolytes: Science and Technology*; Marcel Dekker, Inc.: New York, 1993.

(27) Muthukumar, M. Theory of Viscoelastic Properties of Polyelectrolyte Solutions. *Polymer (Guildf)* **2001**, *42* (13), 5921–5923. https://doi.org/10.1016/S0032-3861(00)00907-1.

(28) Borsali, R.; Vilgis, T. A.; Benmouna, M. Viscosity of Weakly Charged Polyelectrolyte Solutions: The Mode-Mode Coupling Approach. *Macromolecules* **1992**, *25* (20), 5313–5317. https://doi.org/10.1021/ma00046a032.

(29) Odijk, T. Possible Scaling Relations for Semidilute Polyelectrolyte Solutions. *Macromolecules* **1979**, *12* (4), 688–693. https://doi.org/10.1021/ma60070a028.

(30) Wang, L.; Bloomfield, V. A. Cooperative Diffusion of Semidilute Polyelectrolyte Solutions: Analysis by Renormalization Group Theory. *Macromolecules* **1989**, *22* (6), 2742–2746. https://doi.org/10.1021/ma00196a035.

(31) Lopez, C. G.; Richtering, W. Viscosity of Semidilute and Concentrated Nonentangled Flexible Polyelectrolytes in Salt-Free Solution. *J Phys Chem B* **2019**, *123* (26), 5626–5634. https://doi.org/10.1021/acs.jpcb.9b03044.

(32) Dedic, J.; Okur, H. I.; Roke, S. Polyelectrolytes Induce Water-Water Correlations That Result in Dramatic Viscosity Changes and Nuclear Quantum Effects. *Sci Adv* **2019**, *5* (12), eaay1443. https://doi.org/10.1126/sciadv.aay1443.

(33) Mintis, D. G.; Dompé, M.; Kamperman, M.; Mavrantzas, V. G. Effect of Polymer Concentration on the Structure and Dynamics of Short Poly(N, N-Dimethylaminoethyl Methacrylate) in Aqueous Solution: A Combined Experimental and Molecular Dynamics Study. *Journal of Physical Chemistry B* **2020**, *124* (1), 240–252. https://doi.org/10.1021/acs.jpcb.9b08966.

(34) Mayoral, E.; Goicochea, A. G. Modeling of Branched Thickening Polymers under Poiseuille Flow Gives Clues as to How to Increase a Solvent's Viscosity. *Journal of Physical Chemistry B* **2021**, *125* (6), 1692–1704. https://doi.org/10.1021/acs.jpcb.0c11087.

(35) Hoogerbrugge, P. J.; Koelman, J. M. V. A. Simulating Microscopic Hydrodynamic Phenomena with Dissipative Particle Dynamics. *Epl* **1992**, *19* (3), 155–160. https://doi.org/10.1209/0295-5075/19/3/001.

(36) Espanol, P.; Warren, P. Statistical Mechanics of Dissipative Particle Dynamics. *Epl* **1995**, *30* (4), 191–196. https://doi.org/10.1209/0295-5075/30/4/001.

(37) Groot, R. D.; Warren, P. B. Dissipative Particle Dynamics: Bridging the Gap between Atomistic and Mesoscopic Simulation. *Journal of Chemical Physics* **1997**, *107* (11), 4423–4435. https://doi.org/10.1063/1.474784.

(38) Pastorino, C.; Goicochea, A. G. Dissipative Particle Dynamics: A Method to Simulate Soft Matter Systems in Equilibrium and Under Flow. In *Environmental Science and Engineering*; Springer Berlin Heidelberg, 2015; pp 51–79. https://doi.org/10.1007/978-3-319-11487-3_3.

(39) Español, P.; Warren, P. B. Perspective: Dissipative Particle Dynamics. *J Chem Phys* **2017**, *146* (15), 150901. https://doi.org/10.1063/1.4979514.

(40) Li, Z.; Bian, X.; Li, X.; Deng, M.; Tang, Y.-H.; Caswell, B.; Karniadakis, G. E. Dissipative Particle Dynamics: Foundation, Evolution, Implementation, and Applications. In *Particles in Flows. Advances in Mathematical Fluid Mechanics*; Bodnár, T., Galdi, G., Nečasová, Š., Eds.; 2017; pp 255–326. https://doi.org/10.1007/978-3-319-60282-0_5.

(41) Pivkin, I. v.; Karniadakis, G. E. A New Method to Impose No-Slip Boundary Conditions in Dissipative Particle Dynamics. *J Comput Phys* **2005**, *207* (1), 114–128. https://doi.org/10.1016/j.jcp.2005.01.006.

(42) Hernández Velázquez, J. D.; Mejía-Rosales, S.; Gama Goicochea, A. Nanorheology of Poly - and Monodispersed Polymer Brushes under Oscillatory Flow as Models of Epithelial Cancerous and Healthy Cell Brushes. *Polymer (Guildf)* **2017**, *129*, 44–56. https://doi.org/10.1016/j.polymer.2017.09.046.

(43) Grest, G. S.; Kremer, K. Molecular Dynamics Simulation for Polymers in the Presence of a Heat Bath. *Phys Rev A (Coll Park)* **1986**, *33* (5), 3628–3631. https://doi.org/10.1103/PhysRevA.33.3628.

(44) Goicochea, A. G.; Romero-Bastida, M.; López-Rendón, R. Dependence of Thermodynamic Properties of Model Systems on Some Dissipative Particle Dynamics Parameters. *Mol Phys* **2007**, *105* (17–18), 2375–2381. https://doi.org/10.1080/00268970701624679.

(45) Hernández-Fragoso, J. S.; Alas, S. D. J.; Goicochea, A. G. Polymer Chains of a Large Persistence Length in a Polymer Brush Require a Lower Force to Be Compressed Than Chains with a Short Persistence Length. *ACS Appl Polym Mater* **2020**, *2* (11), 5006–5013. https://doi.org/10.1021/acsapm.0c00858.

(46) Barcelos, E. I.; Khani, S.; Boromand, A.; Vieira, L. F.; Lee, J. A.; Peet, J.; Naccache, M. F.; Maia, J. Controlling Particle Penetration and Depletion at the Wall Using Dissipative Particle Dynamics. *Comput Phys Commun* **2021**, *258*, 107618. https://doi.org/10.1016/j.cpc.2020.107618.

(47) Chatterjee, A. Modification to Lees-Edwards Periodic Boundary Condition for Dissipative Particle Dynamics Simulation with High Dissipation Rates. *Mol Simul* **2007**, *33* (15), 1233–1236. https://doi.org/10.1080/08927020701713894.

(48) Palmer, T. L.; Espås, T. A.; Skartlien, R. Effects of Polymer Adsorption on the Effective Viscosity in Microchannel Flows: Phenomenological Slip Layer Model from Molecular Simulations. *J Dispers Sci Technol* **2019**, *40* (2), 264–275. https://doi.org/10.1080/01932691.2018.1467776.

(49) Goicochea, A. G.; Firoozabadi, A. CO2 Viscosification by Functional Molecules from Mesoscale Simulations. *Journal of Physical Chemistry C* **2019**, *123* (48), 29461–29467. https://doi.org/10.1021/acs.jpcc.9b08589.

(50) M. P. Allen; D. J. Tildesley. *Computer Simulations of Liquids*; Oxford University Press: New York, 1987.

(51) Alarcón, F.; Pérez-Hernández, G.; Pérez, E.; Gama Goicochea, A. Coarse-Grained Simulations of the Salt Dependence of the Radius of Gyration of Polyelectrolytes as Models for Biomolecules in Aqueous Solution. *European Biophysics Journal* **2013**, *42* (9), 661–672. https://doi.org/10.1007/s00249-013-0915-z.

(52) González-Melchor, M.; Mayoral, E.; Velázquez, M. E.; Alejandre, J. Electrostatic Interactions in Dissipative Particle Dynamics Using the Ewald Sums. *J Chem Phys* **2006**, *125* (22), 224107. https://doi.org/10.1063/1.2400223.

(53) Ibergay, C.; Malfreyt, P.; Tildesley, D. J. Electrostatic Interactions in Dissipative Particle Dynamics: Toward a Mesoscale Modeling of the Polyelectrolyte Brushes. *J Chem Theory Comput* **2009**, *5* (12), 3245–3259. https://doi.org/10.1021/ct900296s.

(54) Procházka, K.; Limpouchová, Z.; Štěpánek, M.; Šindelka, K.; Lísal, M. DPD Modelling of the Self- and Co-Assembly of Polymers and Polyelectrolytes in Aqueous Media: Impact on Polymer Science. *Polymers (Basel)* **2022**, *14* (3), 404. https://doi.org/10.3390/polym14030404.

(55) Alarcón, F.; Pérez, E.; Gama Goicochea, A. Dissipative Particle Dynamics Simulations of Weak Polyelectrolyte Adsorption on Charged and Neutral Surfaces as a Function of the Degree of Ionization. *Soft Matter* **2013**, *9* (14), 3777–3788. https://doi.org/10.1039/c2sm27332b.

(56) Krause, W. E.; Tan, J. S.; Colby, R. H. Semidilute Solution Rheology of Polyelectrolytes with No Added Salt. *J Polym Sci B Polym Phys* **1999**, *37* (24), 3429–3437. https://doi.org/10.1002/(SICI)1099-0488(19991215)37:24<3429::AID-POLB5>3.0.CO;2-E.

(57) Boris, D. C.; Colby, R. H. Rheology of Sulfonated Polystyrene Solutions. *Macromolecules* **1998**, *31* (17), 5746–5755. https://doi.org/10.1021/ma971884i.

(58) Dou, S.; Colby, R. H. Charge Density Effects in Salt-Free Polyelectrolyte Solution Rheology. *J Polym Sci B Polym Phys* **2006**, *44* (14), 2001–2013. https://doi.org/10.1002/polb.20853.

(59) Prini, R. F.; Lagos, A. E. Tracer Diffusion, Electrical Conductivity, and Viscosity of Aqueous Solutions of Polystyrenesulfonates. *J Polym Sci A* **1964**, *2* (6), 2917–2928. https://doi.org/10.1002/pol.1964.100020640.

(60) Yamaguchi, M.; Wakutsu, M.; Takahashi, Y.; Noda, I. Viscoelastic Properties of Polyelectrolyte Solutions. 2. Steady-State Compliance. *Macromolecules* **1992**, *25* (1), 475–478. https://doi.org/10.1021/ma00027a074.